
\input epsf.tex


\def\ifundefined#1{\expandafter\ifx\csname
#1\endcsname\relax}

\newcount\eqnumber \eqnumber=0
\def\beq{ \global\advance\eqnumber by 1 $$ }
\def\eeq{ \eqno(\the\eqnumber)$$ }
\def\label#1{\ifundefined{#1}
\expandafter\xdef\csname #1\endcsname{\the\eqnumber}
\else\message{label #1 already in use}\fi}
\def\(#1){(\csname #1\endcsname)}

\newcount\refno \refno=0
\def\[#1]{\ifundefined{#1}\advance\refno by 1\expandafter\xdef\csname
#1\endcsname{\the\refno}\fi[\csname #1\endcsname]}
\def\refis[#1]{\item{\csname #1\endcsname.}}


\font\titlething=cmbx10 scaled\magstep2
\font\headingthing=cmbx10 scaled\magstep1
\baselineskip=18pt
\magnification=1100


\def\tr{\;{\rm tr}\;}
\def\intall{\int_{0\leq |k| \leq 1} {{d^d k}\over{(2\pi)^d}}}
\def\intslow{\int_{0\leq |k| \leq 1/2} {{d^d k}\over{(2\pi)^d}}}
\def\intfast{\int_{1/2 < |k| \leq 1} {{d^d k}\over{(2\pi)^d}}}
\def\intintall{\int_{0\leq |k_{1,2,3}| \leq 1}
{{d^d k_1 d^d k_2 d^d k_3}\over{(2\pi)^{3d}}}}
\def\intx{\int d^d x}
\def\intxy{\int d^d x d^d y}
\def\intxyz{\int d^d x d^d y d^d z}
\def\intyz{\int d^d y d^d z}
\def\intxyzw{\int d^d x d^d y d^d z d^d w}
\def\pb{\bar\Phi}
\def\lcol{L_{\rm col.}^{\rm max}}
\def\gur{{{c_d u_0}\over{(1+r_0)^2}}}
\def\fr{f_{r_0}}
\def\fu{f_{u_0}}
\def\fq{f_{q^2}}

\vskip.5in
\noindent {\tt hep-th/9504013}

\noindent G\"oteborg ITP 95-99
\vskip.5in
\centerline{{\titlething ON THE LARGE N LIMIT OF 3D}}
\bigskip
\centerline{{\titlething  AND 4D HERMITIAN MATRIX MODELS}}
\vskip 1in

\centerline{\headingthing G. Ferretti}
\vskip.5cm
\centerline{\sl Institute of Theoretical Physics}
\centerline{\sl  Chalmers University of Technology}
\centerline{\sl S-41296 G\"oteborg, Sweden}
\centerline{\tt ferretti@fy.chalmers.se}
\vskip 1in
\centerline{\headingthing Abstract}

The large $N$ limit of the hermitian matrix model in three
and four Euclidean space-time dimensions
is studied with the help of the approximate Renormalization Group recursion
formula. The planar graphs contributing to wave function, mass and coupling
constant renormalization are identified and summed in this
approximation. In four dimensions the
model fails to have an interacting continuum limit, but in three dimensions
there is a non trivial fixed point for the approximate RG relations.
The critical exponents of the three dimensional model at this fixed point
are $\nu = 0.665069$ and $\eta=0.19882$. The existence (or non existence)
of the fixed point and the critical exponents display a fairly high degree
of universality since they do not seem to depend on the specific
(non universal) assumptions made in the approximation.

\vfill\eject

\bigskip
\centerline{\headingthing 1) Introduction.}
\bigskip

Since 't Hooft's original proposal \[thooft1], the large $N$ limit has
been regarded as one of the most promising attempts to understand strongly
coupled gauge theories. The observation that a field theory simplifies in the
presence of a large symmetry group is, in fact, more general
and has led, among other things, to an understanding
of many linear and non linear $\sigma$
models in dimension\footnote{$^{1)}$}{Throughout this paper,
the letter $d$ refers
to the dimension of (Euclidean) space-time.}
$2\leq d < 4$ \[review1], to the exact solution of matrix
models in dimension $d\leq 1$ \[review2] and to the solution of
$d=2$ QCD \[thooft2].

The great simplification arising in
the study of $\sigma$ models is the vector-like nature of the fields, i.e.
the fact that the number of degrees of freedom grows like $N$. This allows one
to reduce the sum of all the leading "cactus" diagrams to the sum of a finite
number of effective graphs, providing in this way a tractable perturbative
expansion. The leading term in the expansion already captures the relevant
features of the various models, including spontaneous symmetry breaking (or
the lack of it in $d=2$), confinement, dynamical generation of relevant
operators, etc...

The situation is more complicated in the case of matrix models, of which QCD
is the prime example. There, the number of degrees of freedom grows like
$N^2$ and one is faced with the arduous task of summing the leading "planar"
diagrams. This problem has been solved in \[brezin] for the single matrix
model ($d=0$), and for the quantum mechanical anharmonic oscillator
($d=1$), with its remarkable equivalence to a free Fermi gas. These low
dimensional models have been shown to describe $d=2$ quantum gravity and
non critical string theory \[review2].

Alas, progress has been much slower in the study of higher dimensional
matrix models. The factorization properties of the expectation values of
observables \[migdal] has led to the conjecture \[witten]
that the leading order in
$1/N$ should be dominated by a single field configuration,
the "master field".
Very recently the precise meaning of the master field has been clarified
using non commutative probability theory \[master]. This recent development
is very exciting but is also somewhat worrisome: the connection between
the master field and the knowledge of the connected Green functions is so
explicit that it might signify that finding the master field is just as
hard as exactly solving the large $N$ theory without it! It seems likely that
one would have to develop approximations to this construction, treating the
large $N$ theory as an "exact" theory in its own right.

In this paper, we develop an approximation to $d=3$ and $d=4$ hermitian
matrix models with quartic potential using one of the
oldest tools in the non perturbative formulation of quantum field theory:
the approximate renormalization group (RG) recursion formula \[wilson].
We make
some modification to this technique to make it more suitable for the
problem at hand: on the one hand, we relax the restriction of no
wave function renormalization \[golner] to allow for a non trivial anomalous
dimension $\eta$ and, on the other hand, we restrict our
attention to the renormalization of the quartic term in the interaction.
The first change is crucial because, to leading order in $1/N$, there are
already an infinite number of diagrams contributing to $\eta$;
neglecting them would ruin the structure of the RG
flow. The second assumption is not crucial, but it simplifies the analysis
by allowing one to perform explicitly the integrals in the recursion formula
using the results from the single $d=0$ matrix model and to reduce the
single recursion step from an integral equation to an algebraic one.

The main results of this investigation are as follows. In $d=3$ the model
has a non trivial fixed
point in addition to the Gaussian one. The large $N$ critical exponents
at this point are computed to be $\eta=0.19882$ and $\nu=0.66517$.
In $d=4$ the model fails to have a non Gaussian large $N$ limit;
its continuum limit is a free theory. This last result
may very well be due to the roughness of the approximation and should
not discourage us to study the $d=4$ theory further.

The paper is organized as follows. In section two, we present the
model and qualitatively describe its renormalization
in the large $N$ limit. In section three, we give a sketch of a perturbative
calculation to third order in the coupling constant. This section
is not needed for the further calculations which are intrinsically
non-perturbative and it is only given for comparison.
In section four, we derive the approximate
recursion formula. All the assumptions that go into its
derivation are spelled out and explicit expressions for wave function, mass
and coupling constant renormalization are obtained in the large $N$ limit.
In section five, we study the
RG equation derived, establish the existence of
a non Gaussian fixed point for $d=3$ and compute the critical exponents.
We also comment on the non existence of a non Gaussian fixed point in
$d=4$. All the results for the
$d=0$ hermitian model \[brezin] of direct relevance for this paper are
collected in the appendix.

\bigskip
\centerline{\headingthing 2) The hermitian matrix model and its RG.}
\bigskip

The model we study in this paper is the $d=3,\; 4$ hermitian
matrix model with quartic interaction. The field variable is a
$N\times N$ hermitian matrix $\Phi(x) = \Phi(x)^\dagger$. In the spirit of
Wilson's RG, we assume the presence of an effective cut-off $\Lambda_0$
in momentum space. Choosing the mass unit to be
$\Lambda_0$ itself, allows us to
set $\Lambda_0=1$ and represent all quantities in dimensionless units. In
this way we write the Fourier transform of the field as
\beq
    \Phi(x) = \intall e^{i x\cdot k} \Phi(k).
\eeq
The Euclidean action for $\Phi$,  reads
\beq\eqalign{
    S[\Phi]&={1\over 2}\intx\tr\bigg((\nabla\Phi)^2 + r_0 \Phi^2\bigg) +
             {u_0\over N}\intx\tr \Phi^4\cr
           &={1\over 2}\intall(k^2 + r_0)\tr\bigg(\Phi(k)\Phi(-k)\bigg)\cr
           &+{u_0\over N}\intintall\tr\bigg(\Phi(k_1)\Phi(k_2)\Phi(k_3)
             \Phi(-k_1-k_2-k_3)\bigg).} \label{action}
\eeq

It is well known that, in the $1/N$ expansion, the leading term for the free
energy is of order $O(N^2)$. Also, the correct normalization for singlet
operators, in order for them to have a finite
vacuum expectation value, is as follows:
a factor of $1/\sqrt{N}$ for every power of the field and a factor of
$1/N$ for every trace; e.g.:
\beq
        {1\over{N^{1+k/2}}}\tr\Phi^k; \quad
        {1\over{N^{2+(k+l)/2}}}\tr\Phi^k\tr\Phi^l\quad\cdots
\eeq

Action \(action) is, of course, just a truncation of the most general
action in theory space consistent with the symmetries of the problem
(throughout the paper we deal with a parity preserving theory). A generic
term in the most general action would contain products of traces of powers
of the field multiplied by (short ranged) functions of their momenta and
by the appropriate powers of $1/N$:
\beq
     \approx{1\over{N^p}}\int {{d^d k_1}\over{(2\pi)^d}}\cdots u(k_1\cdots)
            \tr\bigg(\Phi(k_1)\Phi(k_2)\cdots\bigg)
            \tr\bigg(\Phi(k_i)\Phi(k_{i+1})\cdots\bigg)\cdots.
\eeq

That \(action) is a consistent truncation can only be checked at the end.
However, it should at least seem reasonable because it includes the three
most relevant operators at the Gaussian point.
Actually, there is a fourth such operator:
$(v_0/N^2) \intx \big(\tr\Phi^2\big)^2$.
One might be tempted simply to dismiss such an operator as trivial because it
can be removed by introducing an auxiliary singlet field $\lambda$ to the
action
$ - (N^2/v_0)\intx \big(\lambda - (v_0/N^2) \tr(\Phi^2)\big)^2$, or
because it only affects the
correlation functions of the theory through the insertion of vacuum diagrams
via
a contact interaction. However, this is not quite enough; if such a term was
generated by the RG it would affect mass renormalization and its effect would
have to be taken into account.

At the cost of being pedantic, we begin by showing that such a
term in not generated by applying the RG transformation on \(action) and
therefore it may be excluded a priori. This boils down to showing that actions
of the form \(action) are mapped into themselves by the RG, modulo terms of
higher
Gaussian dimension. There might be some confusion on this point since
it is well known that the usual exponential relations between the generating
function of the connected and disconnected Green functions,
as well as the Legendre transformation between the generating function of
the connected Green function and the one particle irreducible vertices,
are not valid in the large $N$ limit. Why then should the Wilsonian action
re-exponentiate on itself without generating terms non linear in the trace?
The solution to this puzzle is that there are leading Green functions in
$1/N$ that are not planar (fig. 1) and, while they do not appear in the
generating function, they do contribute to the Wilsonian action.

\vskip.2in
\epsfxsize=4truein
\epsfbox{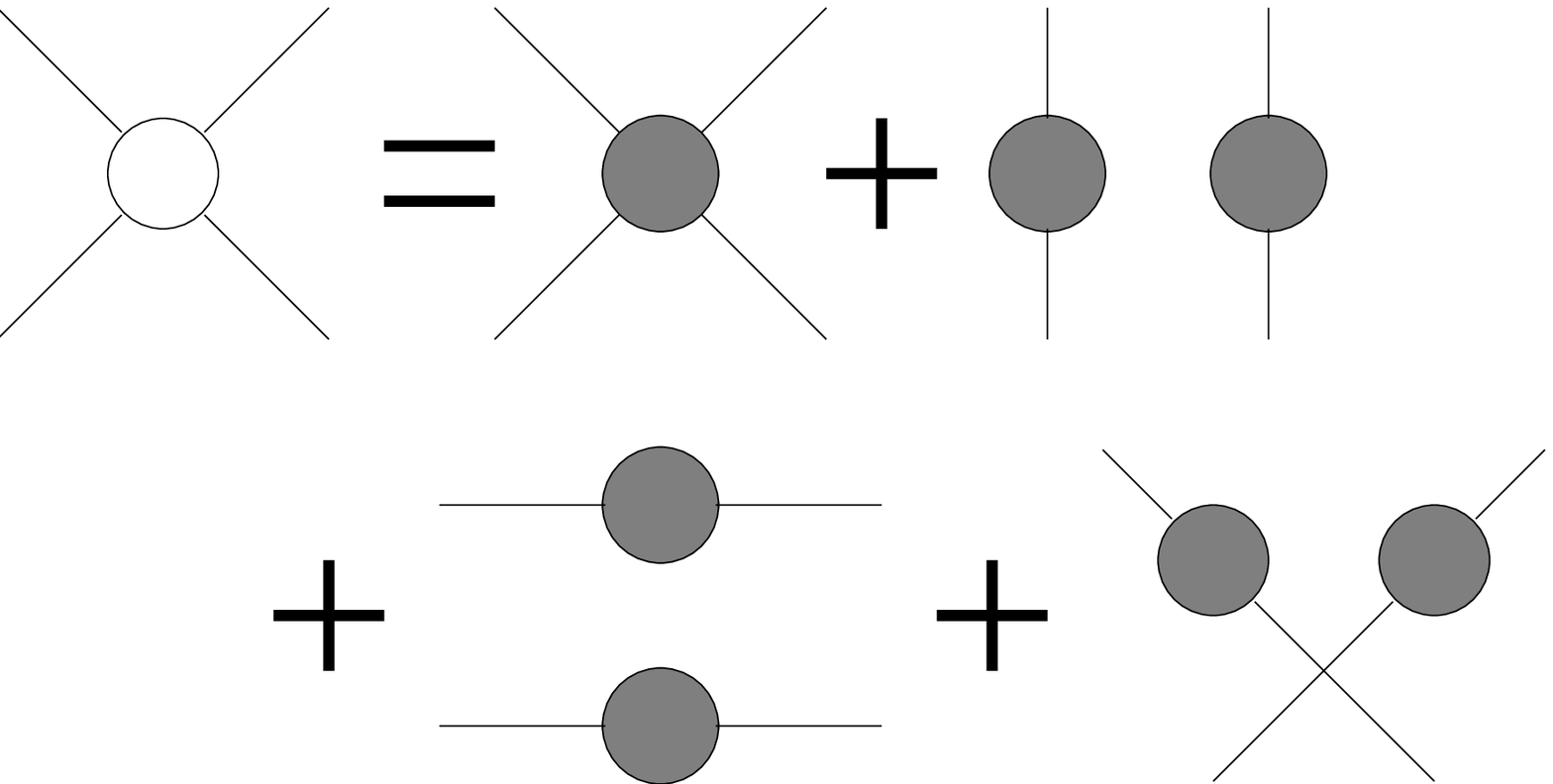}
\vskip.2in
\noindent{{\bf Fig 1.)} The usual decomposition of the four point function into
its connected components. The last contribution is not "planar" but should be
included when computing the renormalized action.}
\vskip.2in

Let us address this and other related issues by carefully
carrying on the first few steps in the renormalization
programme. We must integrate out the higher momenta from the action
and obtain a new action that, when rescaled, will yield information
about the RG flow. Let us then begin by splitting the field variable
into slow and fast components
$\Phi= \pb + \phi$ with momenta smaller and larger than $1/2$ respectively:
\beq
    \pb(x) = \intslow e^{i x\cdot k} \pb(k)
    \quad\hbox{and}\quad
    \phi(x) = \intfast e^{i x\cdot k} \phi(k).
\eeq
Action \(action) decomposes into
\beq
    S[\Phi] = S[\pb] + \sigma[\pb, \phi] + S[\phi], \label{splitaction}
\eeq
where $S[\pb]$ and $S[\phi]$ are the same as \(action) with the appropriate
restriction on the momenta and
\beq
    \sigma[\pb, \phi] = {u_0\over N}\intx\tr\bigg(4\phi\pb^3 +
     4\phi^2\pb^2 + 2(\phi\pb)^2 + 4\phi^3\pb\bigg).
\eeq
(Of course, the quadratic term factorizes by momentum conservation).
Let us denote, for any local functional $F[\pb, \phi]$ of the fields,
\beq
    \bigg<F\bigg> = \lim_{N\to\infty}{{\int{\cal D}\phi e^{-S[\phi]} F}
    \over{\int{\cal D}\phi e^{-S[\phi]}}}. \label{vevdef}
\eeq
The denominator in \(vevdef) is such that $<1>=1$. It removes all
vacuum diagrams just as in the ordinary case. The relation between Green
functions with and without vacuum diagrams is the familiar one because every
leading vacuum diagram is planar and vice versa.

Expanding $\exp(-\sigma)$ to forth order in $\pb$ and keeping only those terms
that are allowed by parity one obtains
\beq\eqalign{
    \Bigg< e^{-\sigma} \Bigg>=&\Bigg< 1-{{u_0}\over{N}}\intx\bigg(
    4\tr\phi^2\pb^2(x) + 2\tr(\phi\pb)^2(x)\bigg)\cr
    &+{1\over2}{{u_0^2}\over{N^2}}\intxy\bigg(
    32\tr\phi\pb^3(x)\tr\phi^3\pb(y)+16\tr\phi^3\pb(x)\tr\phi^3\pb(y)\cr
    &+16\tr\phi^2\pb^2(x)\tr\phi^2\pb^2(y)+16\tr\phi^2\pb^2(x)
    \tr(\phi\pb)^2(y)+4\tr(\phi\pb)^2(x)\tr(\phi\pb)^2(y)\bigg)\cr
    &-{1\over6}{{u_0^3}\over{N^3}}
    \intxyz\bigg(192\tr\phi^2\pb^2(x)\tr\phi^3\pb(y)\tr\phi^3\pb(z)\cr
    &+96\tr(\phi\pb)^2(x)\tr\phi^3\pb(y)\tr\phi^3\pb(z)\bigg)\cr
    &+{1\over{24}}{{u_0^4}\over{N^4}}\intxyzw\bigg(256\tr\phi^3\pb(x)
    \tr\phi^3\pb(y)\tr\phi^3\pb(z)
    \tr\phi^3\pb(w)\bigg) \Bigg>.}\label{expsigma}
\eeq
(This is not a perturbative expansion; each term contains
diagrams with an arbitrary number of $\phi^4$ vertices.)

The first thing to notice is that none of the four terms containing
the operator $\tr(\phi\pb)^2$ in \(expsigma) contributes to the expression.
Terms like those in fig. 2.a yield a contribution proportional to $(1/N)
(\tr\pb)^2$. This term would be leading ($O(N^2)$)
if $\tr\pb$ was allowed to pick up a vacuum expectation value, but it is
subleading in a parity invariant theory:
\beq\eqalign{
        \bigg<{1\over N}(\tr\pb)^2\bigg> &= N^2\times\bigg<{1\over
        {N^3}}(\tr\pb)^2\bigg> = N^2\times\Bigg(\bigg<{1\over
        N^{3/2}}\tr\pb\bigg>^2 + O(N^{-2})\Bigg)\cr &=
        N^2\times\bigg(0 + O(N^{-2})\bigg)= O(1).\cr}
\eeq
Terms like those in fig. 2.b "would be" leading but they vanish identically
because of momentum conservation. In fact, all diagrams containing a
sub-diagram with one $\pb$ leg and one $\phi$ leg (fig. 3) vanish identically
by momentum conservation. This implies that the term $\tr\phi\pb^3(x)
\tr\phi^3\pb(y)$ in \(expsigma) (fig. 2.c) also vanishes identically.
Finally, notice that a term like the one in fig. 2.d is planar but it is
subleading because
the fields $\pb$ are not connected at the outside of the diagram.

\vskip.2in
\epsfxsize=6truein
\epsfbox{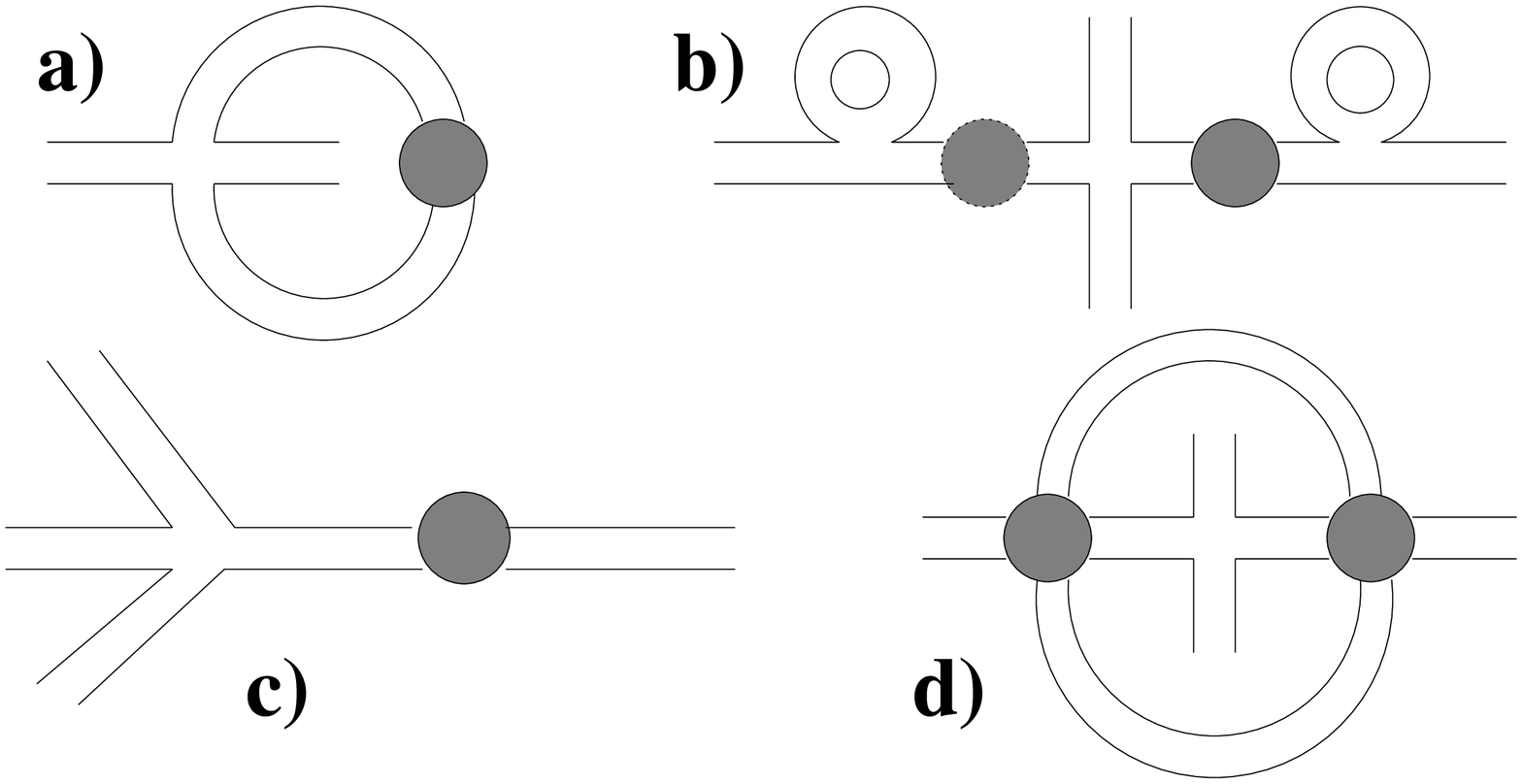}
\vskip.2in
\noindent{{\bf Fig 2.)} Various contributions to the renormalized action in
t' Hooft's double line notation. All external lines represent the slow field
$\pb$ and all the internal lines the fast field $\phi$.
{\bf 2.a)} is subleading if the trace
does not pick up a vacuum expectation value. {\bf 2.b)} and {\bf 2.c)} are
identically zero by momentum conservation. {\bf 2.d)} is planar but
subleading.}
\bigskip
\epsfxsize=3truein
\epsfbox{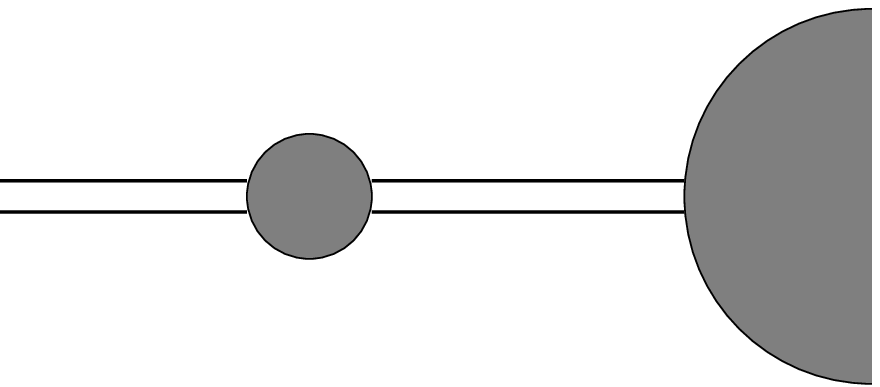}
\vskip.2in
\noindent{{\bf Fig 3.)} Same notation as in fig {\bf 2}.
All diagrams containing this sub-diagram vanish.}
\vfill\eject

Considering all the remaining terms and performing the decomposition
into connected components one obtains:
\beq\eqalign{
    \Bigg< e^{-\sigma} \Bigg>=&1-{{4 u_0}\over{N}}\intx
    \Bigg<\tr\phi^2\pb^2(x) \Bigg>_{\rm conn.} +{{8u_0^2}\over{N^2}}\intxy
    \Bigg<\tr\phi^3\pb(x)\tr\phi^3\pb(y)\Bigg>_{\rm conn.}\cr
    +&{{8u_0^2}\over{N^2}}
    \Bigg(\intx\Bigg<\tr\phi^2\pb^2(x)\Bigg>_{\rm conn.}\Bigg)^2+
    {{8u_0^2}\over{N^2}}\intxy\Bigg<\tr\phi^2\pb^2(x)
    \tr\phi^2\pb^2(y)\Bigg>_{\rm conn.}\cr -&{{32 u_0^3}\over{N^3}}
    \intx\Bigg<\tr\phi^2\pb^2(x)\Bigg>_{\rm conn.}\intyz\Bigg<\tr\phi^3\pb(y)
    \tr\phi^3\pb(z)\Bigg>_{\rm conn.}\cr -&{{32 u_0^3}\over{N^3}}\intxyz
    \Bigg<\tr\phi^2\pb^2(x)\tr\phi^3\pb(y)\tr\phi^3\pb(z)\Bigg>_{\rm conn.}\cr
    +&{{{\bf 32}u_0^4}\over{N^4}}\Bigg(\intxy\Bigg<\tr\phi^3\pb(x)
    \tr\phi^3\pb(y)\Bigg>_{\rm conn.}\Bigg)^2\cr
    +&{{32u_0^4}\over{3 N^4}}\intxyzw
    \Bigg<\tr\phi^3\pb(x)\tr\phi^3\pb(y)\tr\phi^3\pb(z)\tr\phi^3\pb(w)
    \Bigg>_{\rm conn.}.\cr} \label{expsigmanew}
\eeq
The only relevant point is that the next to last quantity
$\big(\intxy\big<\tr\phi^3\pb(x)
\phi^3\pb(y)\big>_{\rm conn.}\big)^2$ is multiplied by the factor
${\bf 32}={\bf 3}\times 256/24$, in other words we had to use all three
disconnected Green functions in fig. 1 since we were taking the
vacuum expectation value of four singlets ($\tr\phi^3\pb$). This makes
it possible to re-exponentiate \(expsigmanew) without generating terms
non linear in the trace. If we denote by $\tilde S[\pb]$ the exponential:
\beq
        e^{- \tilde S[\pb]} = \Bigg<e^{-\sigma[\pb, \phi]}\Bigg>,
\eeq
then
\beq\eqalign{
    \tilde S[\pb] =& {{4 u_0}\over{N}}\intx
    \Bigg<\tr\phi^2\pb^2(x) \Bigg>_{\rm conn.}- {{8u_0^2}\over{N^2}}\intxy
    \Bigg<\tr\phi^3\pb(x)\tr\phi^3\pb(y)\Bigg>_{\rm conn.}\cr
    -&{{8u_0^2}\over{N^2}}\intxy\Bigg<\tr\phi^2\pb^2(x)
    \tr\phi^2\pb^2(y)\Bigg>_{\rm conn.}\cr
    +&{{32 u_0^3}\over{N^3}}\intxyz
    \Bigg<\tr\phi^2\pb^2(x)\tr\phi^3\pb(y)\tr\phi^3\pb(z)\Bigg>_{\rm conn.}\cr
    -&{{32u_0^4}\over{3 N^4}}\intxyzw
    \Bigg<\tr\phi^3\pb(x)\tr\phi^3\pb(y)\tr\phi^3\pb(z)\tr\phi^3\pb(w)
    \Bigg>_{\rm conn.}.\cr} \label{stilde}
\eeq

So far, our discussion has been completely general and no approximation has
been made, except for the truncation to operators containing
no more than four fields $\pb$. Of course, the identification and evaluation
of the diagrams contributing to \(stilde)
is still a formidable problem and we must make further
approximations; these will be made in section four. Before that, in the next
section, we present the sketch of a perturbative calculation as a warm up.

\bigskip
\centerline{\headingthing 3) A perturbative calculation.}
\bigskip

A perturbative expansion to $(O(u_0^3))$
around the Gaussian point and to leading
order in $1/N$ results in the diagrammatic expansion for $\tilde S[\pb]$
given in fig. 4.

\vskip.2in
\epsfysize=6truein
\epsfbox{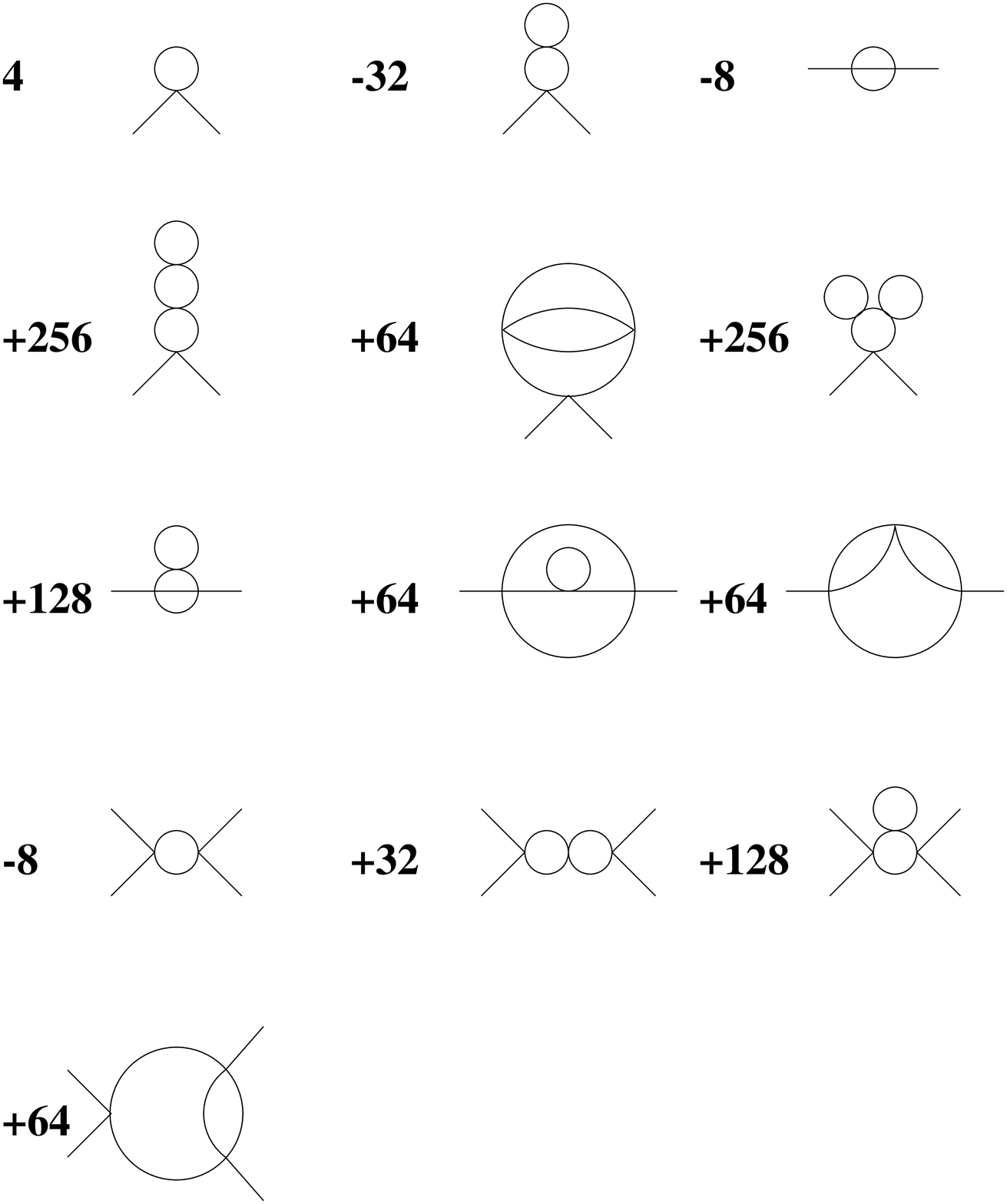}
\vskip.2in
\noindent{{\bf Fig 4.)} Perturbative expansion of the renormalized action
in single line notation. The external lines represent the field $\pb$ and
the internal ones the field $\phi$.}
\vskip.2in

All diagrams that are forbidden by momentum conservation
have been omitted. The symmetry factors are all powers of $2$ and
this is typical of planar expansions. One can compare these factors with those
in \[brezin], keeping in mind that there is an overall factor $2$ of
difference in the two-point function and an overall factor $4$ in the
four-point function. This is so because, here, the external legs are attached
to
the slow field $\pb$ acting as a background field. One must also remember
the explicit factor of $1/N$ in front of the four-point functions.

Just as an example, the first diagram in fig. 4. corresponds to the expression
\beq
        4 u_0 \intx\tr\big(\pb^2(x)\big)G(x,x); \label{firstterm}
\eeq
whereas, say, the tenth diagram corresponds to
\beq
       -{{8 u_0^2}\over{N}}\intxy\tr\big(\pb^2(x)\pb^2(y)\big)G(x,y)^2.
       \label{fourthterm}
\eeq
In \(firstterm) and \(fourthterm), the quantity $G(x,y)$ is given by the tree
level propagator for $\phi$:
\beq\eqalign{
       G(x,y) =&\Big< N^{-2}\tr\phi(x)\phi(y) \Big>_0=\intall e^{ik\cdot(x-y)}
       G(k)\cr G(k)=&\cases{ 0 & for $|k|\leq{1\over2}$\cr &\cr
                 {1\over{k^2 + r_0}} & for ${1\over2}<|k|\leq1$\cr}\cr}
\eeq
The symbol $\bigg<\cdots\bigg>_0$ is the Gaussian average and one can also
write, for each component of the matrix,
\beq
        \bigg<\phi_\beta^\alpha(x)\phi_\delta^\gamma(y)\bigg>_0 =
        \delta^\alpha_\delta \delta_\beta^\gamma G(x,y) + O({1\over N}).
\eeq

One could then perform the integrals and obtain a perturbative expression
for $\tilde S$, but this will not be done explicitly because
the main purpose of this section was to present the relevant diagrams
with their relative weights for later comparison. Let us therefore go
on to the next section and present those approximations that allow one
to obtain non perturbative results.

\bigskip
\centerline{\headingthing 4) The approximate RG formula in the large N limit.}
\bigskip

We must now make some different approximations in order to compute \(stilde)
without using perturbation theory. The problem is, of course, that in the
expression for $\tilde S$ we want to keep the contribution
of integrals of arbitrary higher number
of loops and we do not know how to do it exactly. The first thing we do
is to set all incoming momenta (the momenta of
the slow field $\pb$) to zero, in order to obtain the "ultra local"
contributions
to $\tilde S$. This approximation is equivalent to the truncation we have
chosen for \(action) and it is not an independent assumption,
but it should {\it only} be applied to those terms that
are already ultra local in \(action),
namely, mass term and interaction. If we applied
it to the kinetic term we would rule out the possibility of
anomalous dimension for the field. This would be too strong since, in the
large $N$ limit, there is an infinite number of planar diagrams contributing
to wave function renormalization. Let us therefore start with the mass and
the interaction terms, leaving the analysis of the kinetic term for later.

In the spirit of the approximate recursion formula, we now make two further
assumptions:
\item{1.} Replace every propagator (regardless of its momentum
dependence) by $1/(1 + r_0)$
\item{2.}JReplace every loop integral by a constant $c_d$

The actual value of the constant $c_d$ is not important. To be specific, we
will assume
\beq
   c_d = \intfast 1 \approx\cases{ 0.014776 & for $d=3$\cr
                         0.002968 & for $d=4$}. \label{cddef}
\eeq

Now we must identify those diagrams that contribute to $\tilde S$ and
estimate the relative powers of $u_0$, $1/(1+r_0)$ and $c_d$ in each
of them. Since all external momenta are set to zero, momentum
conservation implies that not only diagrams containing sub-diagrams like
the one in fig. 3, but all diagrams that can be disconnected by cutting one
internal $\phi$ line vanish, no matter how many external $\pb$ lines are
attached to it. Therefore, only one particle irreducible diagrams contribute,
as one would naively have expected. \footnote{$^{2)}$}
{This observation is not directly relevant
to the calculation of the corrections to $r_0$ and $u_0$ because, due to the
particular nature of quartic interactions, all one particle reducible diagrams
with two or four external legs have a sub-diagram of the kind in fig. 3;
but it should be kept in mind if one wants to compute diagrams with more
than four external legs.}
The great simplification in considering only planar diagrams is that,
with the assumptions above, the sum of these diagrams can be explicitly
evaluated by using the single matrix model (c.f.r. appendix).  None of the
further assumptions usually made in the approximate recursion formula,
(such as forbidding those diagrams with an odd number of internal lines
at some vertex,) need to be made. The situation is
summarized in fig. 5. The diagram in fig. 5.a is exactly forbidden;
the diagram in fig 5.b is forbidden under the assumption of zero incoming
momentum; the diagram in fig. 5.c, that would be forbidden by the ordinary
approximation, is in fact allowed, as it should.

\vskip.2in
\epsfxsize=4truein
\epsfbox{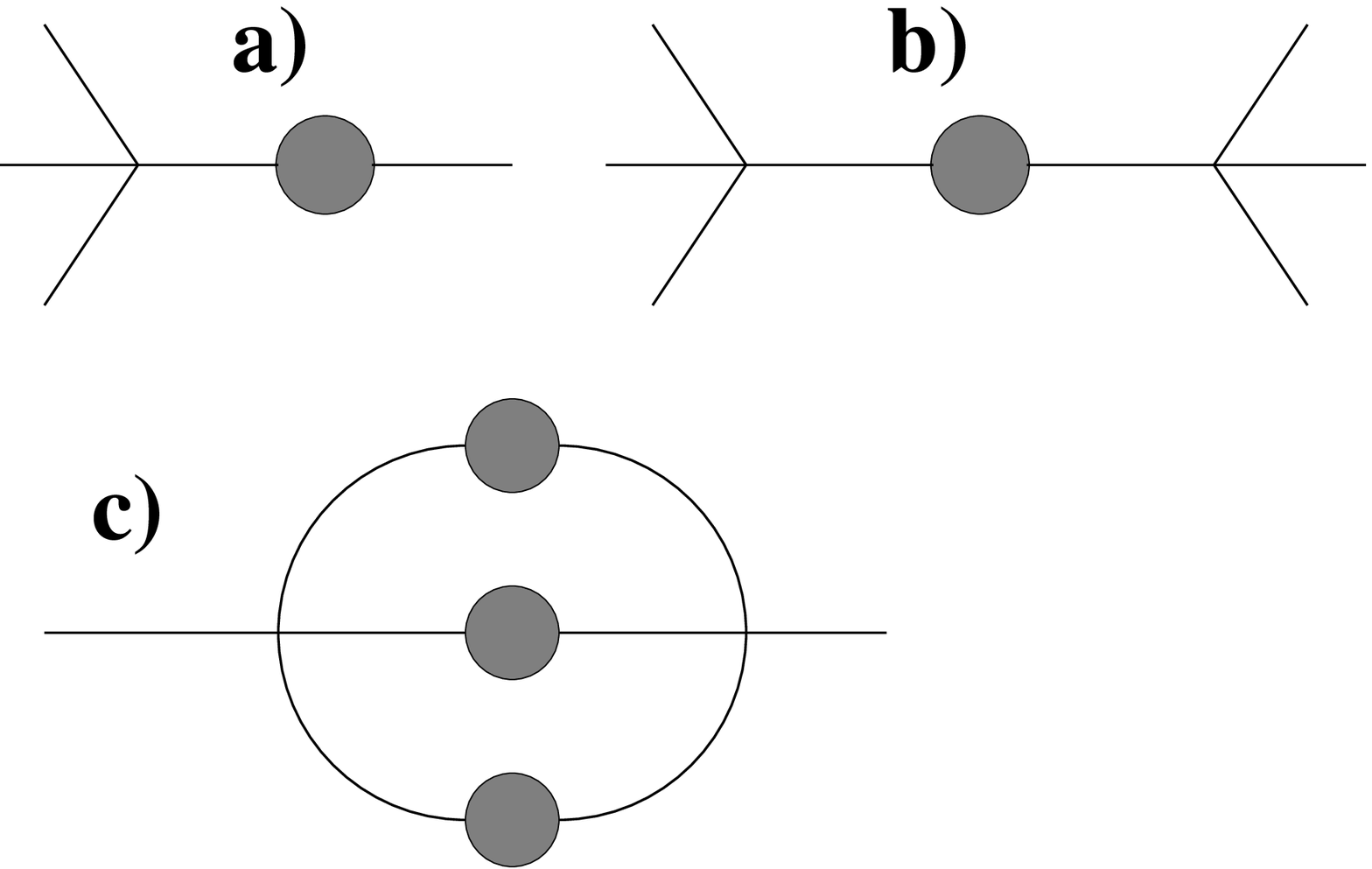}
\vskip.2in
\noindent{{\bf Fig 5.)} Some planar diagrams in single line
notation. {\bf 5.a)}
has already been shown to be identically zero. {\bf 5.b)} vanishes in
the approximations used in this section. {\bf 5.c)} does not vanish and
it is accounted for in our formulas.}
\vskip.2in

To estimate the relative powers of $u_0$, $1/(1+r_0)$ and $c_d$, we make
use of the following topological formulas: for a particular diagram with
$V$ vertices, $E>0$ external $\pb$ legs, $P$ propagators and $L$
{\it momentum} loops, the following two relations apply
\beq
       4V = E + 2P  \quad \hbox{and} \quad V - P + L =1. \label{topological}
\eeq
The reader should not be confused by the second relation in \(topological).
In \(topological), $L$ refers to the number of loop momentum integrals,
not color loops. Among
all diagrams with $L$ momentum loops, the planar limit picks out those with
the maximum number $\lcol$ of color loops. For $E>0$ these two numbers
coincide, whereas for vacuum diagrams $\lcol = L+1$.

For a given number $E$ of external legs, we can solve for $P$ and $L$ as
function of the number of vertices $V$.
We can therefore write the mass term ($E=2$) $\tilde S_{r_0}$ and the
interaction term ($E=4$) $\tilde S_{u_0}$ in $\tilde S$ as follows
\beq\eqalign{
     \tilde S_{r_0}[\pb]&=(1+r_0) \fr\Bigg(\gur\Bigg)\intx\tr\pb^2(x)\cr
     \tilde S_{u_0}[\pb]&={{u_0}\over{N}}
              \fu\Bigg(\gur\Bigg)\intx\tr\pb^4(x),\cr}
\eeq
where $\fr(g)$ and $\fu(g)$ are given in terms of the single matrix model
proper vertices (c.f.r. appendix):
\beq
       \fr(g) = {{\Gamma_2(g) - 1}\over{2}} \quad\hbox{and}\quad
       \fu(g) = {{\Gamma_4(g) - 4g}\over{4g}}.
\eeq
After all that has been said, the proof follows easily. First, we match the
powers of $g$ in the single matrix model with the powers of $u_0$ in
the full quantum theory by dividing by the overall factors of $2$
and $4$ already encountered in perturbation theory.\footnote{$^{3)}$}
{The explicit
factor of $g$ at the denominator in the definition of $\fu$ is there
simply because we have factored out one power of $u_0$ in the definition
of $\tilde S_{u_0}$, it does not represent a singularity at the origin!}
Second, we subtract the tree level terms, already present in $S[\pb]$
\(splitaction) and not to be included in $\tilde S[\pb]$ \(stilde).
Finally, we match the powers of $c_d$ and $1/(1+r_0)$ with the help
of the topological formulas \(topological).

We must now face the problem of wave function renormalization by relaxing
the condition of zero external momenta.
Let $q<<1/2$ be the external momentum of the fields $\pb$. First we
isolate those diagrams that contribute to the
renormalization: they are those with two external
legs connected at two {\it different} points in space-time.
Terms where the two external legs are connected to the same point
do not have any explicit dependence on the external momenta and they
should not be included. The decomposition
of the one particle irreducible two point function into these two kind of terms
is given in fig. 6.

\vskip.2in
\epsfxsize=4truein
\epsfbox{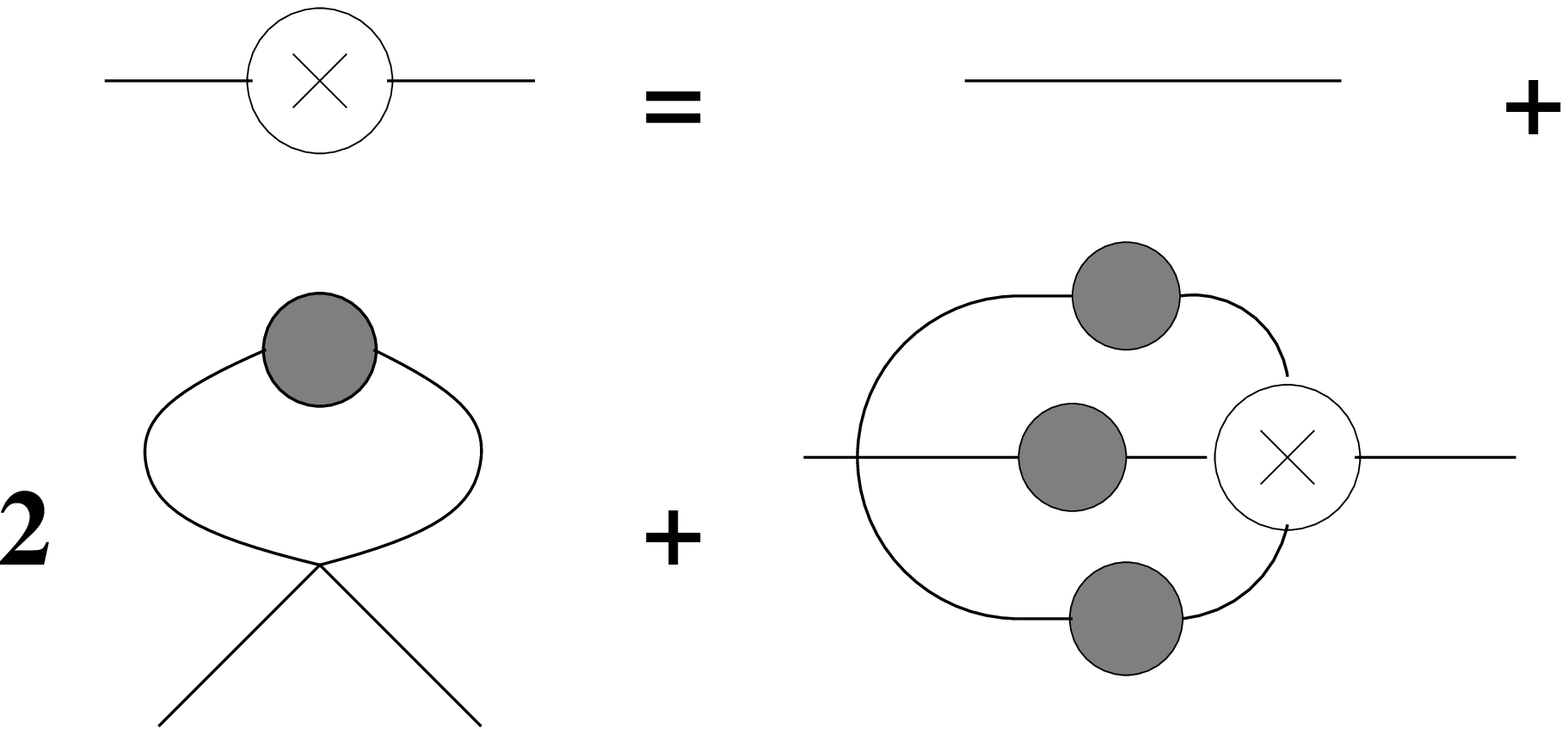}
\vskip.2in
\noindent{{\bf Fig 6.)} A useful decomposition of the vertex $\Gamma_2$.
Solid blobs represent connected Green functions and crossed blobs represent
one particle irreducible vertices. The last term contributes to wave
function renormalization.}
\vskip.2in

Fig. 6 can be checked explicitly for the single matrix model
(c.f.r. appendix):
\beq
        \Gamma_2(g) = 1 + 8g G_2(g) - 4g \Gamma_4(g) G_2^3(g).
        \label{scwhingerdyson}
\eeq
Only the last term in \(scwhingerdyson) contributes to wave function
renormalization. To find such contribution we must take the derivative
with respect to $q^2$ of each contributing diagram and then set $q^2=0$.
Again, this is hard to do exactly because there may be more than one $q$
dependent propagator in a diagram. We overcome this difficulty by
making the further approximation, (justified by dimensional analysis and
also by acting with the derivative on a single propagator):
\beq
        {d\over{dq^2}}\Bigg|_0 \rightarrow -{1\over{1+r_0}}.
\eeq
This allows us to estimate the planar contributions without any effort.
It is an approximation in the same spirit as those made before. It should
correctly capture the qualitative behavior and the order of magnitude
of the correction. By using it, we obtain the last piece of $\tilde S$:
\beq
        \tilde S_{q^2}[\pb]=\fq\Bigg(\gur\Bigg)
        \intx\tr\bigg(\nabla\pb(x)\bigg)^2,
\eeq
where (remembering the $1/2$ normalization factor as before),
\beq
       \fq(g) = 2g\Gamma_4(g)G_2^3(g).
\eeq
Action \(stilde) is therefore specified in terms of three known
functions $\fr$, $\fu$ and $\fq$ of
\beq
         g(u_0, r_0) = \gur. \label{gurdef}
\eeq
These three functions can be written as rational functions of a single
function of $g$ (c.f.r. appendix)
\beq
       a^2(g) ={{\sqrt{1 + 48g} -1}\over{24g}}
\eeq
as follows
\beq\eqalign{
        \fr(g)=&{{(a^2-1)(a^2-3)}\over{2a^2(4-a^2)}}
                \approx 4 g - 40 g^2 + 832 g^3 + \cdots\cr
        \fu(g)=&{{27(5-2a^2)}\over{(4-a^2)^4}} - 1
                \approx -8g + 224 g^2 - 7296 g^3 + \cdots\cr
        \fq(g)=&{{(a^2-1)^2(5-2a^2)}\over{18a^2(4-a^2)}}
                \approx 8 g^2 -256 g^3 + \cdots.\cr} \label{threefunctions}
\eeq
A plot of these three functions is given in fig. 7.
The range in which they are real is $-1/48 \leq g <\infty$; they are analytic
at $g=0$ and their expansion given in \(threefunctions) matches the
perturbative expansion in fig. 4 if one makes the
the same approximations as we have made here.

\vskip.2in
\epsfxsize=4truein
\epsfbox{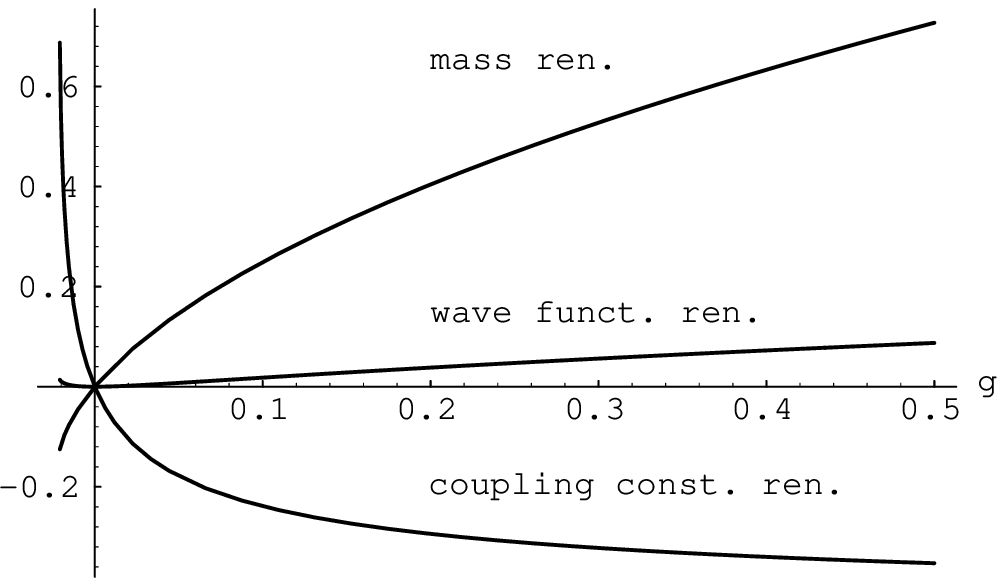}
\vskip.2in
\noindent{{\bf Fig 7.)} A plot of the three functions described in the text.
$\fr$ is the mass renormalization, $\fu$ the coupling constant renormalization
and $\fq$ the wave function renormalization. $\fu$ approaches a finite limit
as $g\to\infty$ whereas $\fr$ and $\fq$ diverge as $\sqrt{g}$. }
\vskip.2in

Let us summarize what has been done so far. We started with the action
$S[\Phi]$ \(action), and after integrating out the fast modes $\phi$
 we obtained a new action for the slow fields $\pb$:
\beq\eqalign{
       & S^\prime[\pb] = S[\pb] + \tilde S[\pb] = {1\over2}\intx\Bigg[1+
       2\fq\Bigg(\gur\Bigg)\Bigg]\tr\bigg(\nabla\pb(x)\bigg)^2\cr +&
       {1\over2}\intx\Bigg[r_0 + 2(1+r_0)\fr\Bigg(\gur\Bigg)\Bigg]\tr
       \pb(x)^2 + {{u_0}\over{N}}\intx
       \Bigg[1+\fu\Bigg(\gur\Bigg)\Bigg]\tr\pb(x)^4.}
\eeq

To obtain the RG equations we have to rescale co-ordinates
$x\to 2x$, ($k\to k/2$) and redefine a new field
\beq
       \Phi(x) = 2^{(d-2)/2}\sqrt{1 + 2\fq\bigg(\gur\bigg)}\pb(2x)
\eeq
in terms of which the action becomes again of the form \(action), but with
a new mass term and a new coupling constant
\beq
       S[\Phi]={1\over 2}\intx\tr\bigg((\nabla\Phi)^2 + r_1 \Phi^2\bigg) +
      {u_1\over N}\intx\tr \Phi^4.
\eeq
This is just the first step of the RG transformation. The same relation
between $r_0,u_0$ and $r_1,u_1$ holds for a generic step between
$r_l,u_l$ and $r_{l+1},u_{l+1}$. We have therefore obtained the RG equations
for the hermitian matrix model in this approximation:
\beq\eqalign{
        r_{l+1}=&4{{r_l + 2(1+r_l)\fr(g)}\over{1+2\fq(g)}}\cr
        u_{l+1}=&2^{4-d} u_l {{1 + \fu(g)}\over{(1+2\fq(g))^2}},\cr}
        \label{rgstep}
\eeq
where $g=g(u_l,r_l)$ is given in \(gurdef).
In the next and final section we investigate these equations to determine
the dynamics of the theory.

\bigskip
\centerline{\headingthing 5) Study of the RG. Fixed point and critical
exponents.}
\bigskip

The iteration of eqs. \(rgstep) determines the RG flow in the two
dimensional space spanned by two co-ordinates $r$ and $u$.
At a fixed point $r^*, u^*$, any further iteration leaves the co-ordinates
invariant, i.e., $r^* = r_l = r_{l+1}$ and $u^*=u_l=u_{l+1}$. Let us also
define, for convenience, the quantity
\beq
    g^*=g(u^*, r^*) = {{c_d u^*}\over{(1+r^*)^2}}. \label{gstardef}
\eeq
At the fixed point, eqs. \(rgstep) then become
\beq\eqalign{
        r^*=&4{{r^* + 2(1+r^*)\fr(g^*)}\over{1+2\fq(g^*)}}\cr
        u^*=&2^{4-d} u^* {{1 + \fu(g^*)}\over{(1+2\fq(g^*))^2}}.\cr}
        \label{rgfixed}
\eeq
By eliminating the solution $u^*=0$ (Gaussian point) from the second eq. in
\(rgfixed), we obtain an algebraic equation is terms of the variable $g^*$
alone:
\beq
     2^{d-4}(1+2\fq(g^*))^2 = 1 + \fu(g^*). \label{gstar}
\eeq
For $d=4$, one can see right away
that the only solution is $g^*=0$. In fact, for
$g>0$ (physical region), it is always $\fq(g)>0$ and $\fu(g)<0$ (fig. 7), thus
preventing \(gstar) from having non trivial solutions\footnote{$^{4)}$}
{One can also check that there are no solutions in the interval
$[-1/48, 0[$}. Within our approximation, this lack of a
non trivial fixed point in four dimensions is rather robust
since it only depends on the relative sign of $\fq$ and $\fu$.
However, it should not be interpreted as a "no go" result because more subtle
approximations might reveal a richer structure.

Things are much better in $d=3$. We can immediately see that eq. \(gstar)
must have a non trivial solution in the physical region by considering the
asymptotic behaviour of the two sides of the equation as $g\to 0$ and
$g\to \infty$. As $g\to 0$, the l.h.s. $\to 1/2$ and the r.h.s. $\to 1$,
whereas
as $g\to\infty$, the r.h.s. approaches a finite value and the l.h.s. blows up.
The two functions must then cross somewhere. Numerically, the only two
solution in the whole interval $[-1/48, +\infty[$ are the Gaussian point and
$g^*\approx 0.408655$. By plugging in this last value
to the first of \(rgfixed) one
obtains $r^* = -0.642902$ and, recalling the definitions of $g^*$ \(gstardef)
and of $c_d$ \(cddef) one also obtains $u^* = 3.52675$.

If one studies the evolution of this fixed point as $d$ goes from $3$ to $4$,
one finds that it approaches the Gaussian one. In this sense, our fixed point
is
"Wilson-Fisher-like". On the other hand, our result does not require using the
$\epsilon$ expansion, and it is a truly large $N$ result because it hinges
on the sum of planar graphs and the consequent asymptotic behavior of the
functions $\fr$, $\fu$ and $\fq$ for large $g$. As a final comment, notice
that, if we neglect wave function renormalization by setting $\fq\equiv 0$, the
fixed point goes away, yet another indication of the different nature of
the calculation.

To compute the large $N$ critical exponent $\nu$ we linearize the RG equations
\(rgstep) near the fixed point. Numerically:
\beq\eqalign{
     r_{l+1} - r^* =& 1.97107 (r_l - r^*) + 0.303152(u_l - u^*)\cr
     u_{l+1} - u^* =& 6.11699 (r_l - r^*) + 0.690315 (u_l - u^*).\cr}
\eeq
The largest eigenvalue $\lambda_{\rm max.}$ of the matrix
\beq
    M=\pmatrix{1.97107 & 0.303152 \cr
               6.11699 & 0.690315 \cr}
\eeq
is $\lambda_{\rm max.}=2.8355$, yielding a critical exponent
\beq
     \nu = {{\log 2}\over{\log \lambda_{\rm max.}}} = 0.665069.
\eeq
As a "check" of universality, one can compute the dependence of $M$ on the
non universal quantity $c_d$ and notice that, while the off diagonal entries
of $M$ depend on $c_d$, the characteristic polynomial (i.e., trace and
determinant) does not; and therefore neither does $\nu$.

To compute the anomalous dimension $\eta$ of the matrix field we simply express
the wave function renormalization as a power of $2$ and take the logarithm:
\beq
    \eta = {{\log (1 + 2\fq(g^*))}\over{\log 2}} = 0.19882.
\eeq
The exponent $\eta$ is also independent on $c_d$.

\bigskip
\centerline{\headingthing Acknowledgements and final remarks.}
\bigskip
I would like to thank Per Salomonson for useful discussions.

The numerical calculations have been performed
with the program Mathematica$^\copyright$ running on a SPARK station.

\vfill\eject

\centerline{\headingthing Appendix: Some useful facts about the single
hermitian matrix model.}
\bigskip

In this appendix we collect, without proof, some basic facts about the single
matrix model that are used throughout the paper. All details can be found in
\[brezin].

Let $\phi$ be an $N\times N$ hermitian matrix.
The action of the $d=0$ hermitian matrix model we are interested in is
\beq
        S^{d=0}[\phi] = \tr\left( {1\over2}\phi^2 + {g\over N}\phi^4\right).
\eeq
For every function $F(\phi)$ we define
\beq
       \bigg<F(\phi)\bigg>^{d=0}
       = \lim_{N\to\infty}{{\int d^{N^2}\phi
        e^{-S^{d=0}[\phi]}F(\phi)}
       \over{\int d^{N^2}\phi e^{-S^{d=0}[\phi]}} }.
\eeq
The Green functions are defined as
\beq
       G_{2p}(g) =\bigg< {1\over{N^{1+p}}}\tr\phi^{2p}\bigg>^{d=0}
       + O({1\over N^2}),
\eeq
where the factors of $1/N$ are chosen so that the leading (planar)
contribution is finite.

The relation between the two and four point Green functions $G_2$ and $G_4$,
the connected ones $C_2$ and $C_4$ and the vertices
$\Gamma_2$ and $\Gamma_4$ is
\beq\eqalign{
       C_2 &= G_2\cr
       C_4 &= G_4 - 2G_2^2\cr
       \Gamma_2 &= C_2^{-1} = G_2^{-1}\cr
       \Gamma_4 &= -C_4 C_2^{-4} = (2G_2^2 - G_4) G_2^{-4}\cr}
\eeq

The $g$ dependence of all planar Green functions and vertex functions
can be written in terms of rational functions of $a^2(g)$ where
\beq
        12ga^4 + a^2 -1 = 0 \quad\hbox{i.e.,}\quad
         a^2(g) ={{\sqrt{1 + 48g} -1}\over{24g}}.
\eeq
In particular, for the Green functions of interest:
\beq\eqalign{
       G_2 &={1\over 3} a^2(4-a^2) \approx  1-8g+144g^2-3456g^3+\cdots\cr
       G_4 &= a^4(3-a^2)\approx 2 -36g +864g^2 -24192g^3+\cdots\cr
       C_2 &= {1\over 3} a^2(4-a^2) \approx  1-8g+144g^2-3456g^3+\cdots\cr
       C_4 &=-{{a^4}\over{9}}(1-a^2)(5-2a^2) \approx -4g +160g^2
       -5760g^3+\cdots\cr
       \Gamma_2 &= {3\over{a^2(4-a^2)}}\approx
                                1+8g-80g^2+1664g^3+\cdots\cr
       \Gamma_4 &= {{9(1-a^2)(5-2a^2)}\over{a^4(4-a^2)^4}}\approx
                                  4g-32g^2+896g^3+\cdots\cr}
\eeq
\vfill\eject

\centerline{\headingthing References. \footnote{$^{5)}$}
{\rm The literature on the large $N$ limit and on the renormalization
group is immense. I have been forced to leave out many relevant papers and
to refer to recent reviews whenever possible.}}
\bigskip

\refis[thooft1] G. 't Hooft, Nucl. Phys. {\bf B72} (1974) 461.
\refis[review1] See papers reprinted in: {\it The large $N$
Expansion in Quantum Field Theory and Statistical Mechanics}, eds. E. Brezin
and S. Wadia, World Scientific (1994).
\refis[review2] For a review see P. Di Francesco, P. Ginsparg and J.
Zinn-Justin, Phys. Rep. {\bf 254} (1995) 1; and references therein.
\refis[thooft2] G. 't Hooft, Nucl. Phys. {\bf B75} (1974) 461.
\refis[brezin]  E. Brezin, C. Itzykson, G. Parisi and J. Zuber,
Comm. Math. Phys.  {\bf 59} 35 (1978).
\refis[migdal] A.  Migdal,  Ann. Phys. {\bf 109} (1977) 365.
\refis[witten]  E.  Witten,  in {\it Recent Developments in Gauge Theories}
eds. G. 't Hooft et. al., Plenum Press, (1980).
\refis[master]
J. Greensite and M.B. Halpern. Nucl. Phys. {\bf B211} (1983) 343;\hfill\break
D.V. Voiculescu, K.J. Dykema and A. Nica
{\it Free Random Variables} AMS , Providence (1992);\hfill\break
M. Douglas, Preprint RU-94-81, {\tt hep-th/9411025} (1994);\hfill\break
R. Gopakumar and D.J. Gross, Preprint PUPT-1520, {\tt hep-th/9411021}(1994).
\refis[wilson]
K.G. Wilson, Phys. Lett. {\bf B4} (1971) 3184;\hfill\break
K.G. Wilson and J. Kogut, Phys. Rep. {\bf 12} (1974) 76;
and references therein.
\refis[golner] G. Golner, Phys. Rev. {\bf B8} (1973) 339.

\bye